\documentclass[prl,twoside,preprintnumbers,superscriptaddress,twocolumn,nofootinbib,nobibnotes]{revtex4}
\usepackage{amsmath,graphicx,slashed,multirow,color,ulem}
\usepackage{graphicx,epsfig,amssymb,bm,slashed,wasysym,multirow}
\usepackage[utf8]{inputenc}
\def \beq{\begin{equation}}
\def \eeq{\end{equation}}
\def \beqa{\begin{eqnarray}}
\def \eeqa{\end{eqnarray}}

\usepackage{ragged2e}
\usepackage{booktabs}

\usepackage{tabularx}
\newcolumntype{C}[1]{>{\centering\arraybackslash}p{#1}}

\usepackage{graphicx}
\usepackage{url}
\usepackage{cancel}
\usepackage{placeins}
\usepackage{bm}
\usepackage{hyperref}
\usepackage{amsmath}
\usepackage{amssymb}
\usepackage{listings}
\usepackage[toc, page]{appendix}
\usepackage{color}

\usepackage{tabularx}
\newcolumntype{C}[1]{>{\centering\arraybackslash}p{#1}}

\def\bea{\begin{eqnarray}}
\def\eea{\end{eqnarray}}

\def\gsim{\mathrel{\rlap{\lower4pt\hbox{\hskip1pt$\sim$}}
    \raise1pt\hbox{$>$}}}         
\def\lsim{\mathrel{\rlap{\lower4pt\hbox{\hskip1pt$\sim$}}
    \raise1pt\hbox{$<$}}}         

\preprint{Edinburgh 2023/22}
\preprint{TIF-UNIMI-2023-28}
\preprint{Nikhef 2023-012}
\preprint{CERN-TH-2023-196}

\newcommand{\be}{\begin{equation}}
\newcommand{\ee}{\end{equation}}
\newcommand{\bi}{\begin{itemize}}
\newcommand{\ei}{\end{itemize}}
\newcommand{\ben}{\begin{enumerate}}
\newcommand{\een}{\end{enumerate}}

\newcommand{\lp}{\left(}
\newcommand{\rp}{\right)}

\begin{document}
\title{The intrinsic charm quark valence distribution  of the proton}

\author{Richard D. Ball}
\affiliation{The Higgs Centre for Theoretical Physics, University of Edinburgh,
  JCMB, KB, Mayfield Rd, Edinburgh EH9 3JZ, Scotland}

\author{Alessandro Candido}
\affiliation{Tif Lab, Dipartimento di Fisica, Universit\`a di Milano and
  INFN, Sezione di Milano, Via Celoria 16, I-20133 Milano, Italy}
\affiliation{CERN, Theoretical Physics Department, CH-1211 Geneva 23, Switzerland}

\author{Juan Cruz-Martinez}
\affiliation{CERN, Theoretical Physics Department, CH-1211 Geneva 23, Switzerland}

\author{Stefano Forte}
\affiliation{Tif Lab, Dipartimento di Fisica, Universit\`a di Milano and
  INFN, Sezione di Milano, Via Celoria 16, I-20133 Milano, Italy}

\author{Tommaso Giani}
\affiliation{Department of Physics and Astronomy, Vrije Universiteit, NL-1081 HV Amsterdam }
\affiliation{Nikhef Theory Group, Science Park 105, 1098 XG Amsterdam, The Netherlands}

\author{Felix Hekhorn}
\affiliation{Tif Lab, Dipartimento di Fisica, Universit\`a di Milano and
  INFN, Sezione di Milano, Via Celoria 16, I-20133 Milano, Italy}
\affiliation{University of Jyv\"askyl\"a, Department of Physics, P.O. Box 35, FI-40014 University of Jyv\"askyl\"a, Finland}
\affiliation{Helsinki Institute of Physics, P.O. Box 64, FI-00014 University of Helsinki, Finland}

\author{Giacomo Magni}
\affiliation{Department of Physics and Astronomy, Vrije Universiteit, NL-1081 HV Amsterdam }
\affiliation{Nikhef Theory Group, Science Park 105, 1098 XG Amsterdam, The Netherlands}

\author{Emanuele R. Nocera}
\affiliation{Dipartimento di Fisica, Universit\`a degli Studi di Torino and
INFN, Sezione di Torino, Via Pietro Giuria 1, I-10125 Torino, Italy}

\author{Juan Rojo}
\affiliation{Department of Physics and Astronomy, Vrije Universiteit, NL-1081 HV Amsterdam }
\affiliation{Nikhef Theory Group, Science Park 105, 1098 XG Amsterdam, The Netherlands}

\author{Roy Stegeman}
\affiliation{The Higgs Centre for Theoretical Physics, University of Edinburgh,
  JCMB, KB, Mayfield Rd, Edinburgh EH9 3JZ, Scotland}

\collaboration{The NNPDF Collaboration}

\date{\today}

\begin{abstract}

  We provide a first quantitative
  indication that the wave function of the proton
  contains unequal distributions of charm quarks and antiquarks, i.e.\ a
  nonvanishing intrinsic valence charm distribution.
  A significant nonvanishing valence component cannot be
  perturbatively generated, hence our results reinforce previous 
  evidence that the proton  contains an intrinsic (i.e.,
  not radiatively generated) charm quark component.
  We establish our
  result through a determination of the  parton distribution functions (PDFs) of 
  charm quarks and antiquarks in the proton.
  We propose two novel experimental probes of
  this intrinsic charm valence component: $D$-meson asymmetries in
  $Z$+$c$-jet production at the LHCb experiment, and 
  flavor-tagged structure functions at the Electron-Ion Collider.
    
\end{abstract}

\maketitle


\noindent
{\bf Introduction.}
The possible existence of charm quarks as intrinsic  constituents
of the proton, on the same footing as the much lighter
up, down, and strange quarks,  has fascinated physicists for more
than four decades~\cite{Brodsky:1980pb,Brodsky:2015fna}. 
Charm quarks and antiquarks are heavier  ($m_c\sim 1.5$ GeV) than the proton itself
($m_p\sim 1$ GeV). They are  copiously pair-produced
through  the perturbative QCD radiation of gluons and light quarks
that generates their scale dependence. An
intrinsic charm (IC) component is the scale-independent result that  is left after subtracting this radiative
contribution. 

A plethora of experimental and theoretical studies have tried to either identify
or reject the presence of IC in the proton~\cite{Jimenez-Delgado:2014zga,Ball:2016neh,Guzzi:2022rca,Hou:2017khm}.
We have recently presented a determination of intrinsic charm
in the proton from a global analysis of parton distribution functions
(PDFs)~\cite{NNPDF:2021uiq,NNPDF:2021njg,Ball:2022qks}.
This study found evidence for IC at the $3\sigma$ level, and was supported by independent constraints from
forward $Z$ production  with charm jets at the LHCb experiment~\cite{LHCb:2021stx}.

In Ref.~\cite{Ball:2022qks} we 
determined the distributions of charm quarks and antiquarks
assuming equality of the intrinsic (scale-independent) charm and
anticharm  PDFs, i.e.\ the vanishing of the charm valence PDF
\begin{equation}
  \label{eq:cmin}
c^-(x,Q^2)=c(x,Q^2)-\bar c(x,Q^2)\, .
\end{equation}
The valence charm PDF $c^-(x,Q^2)$ must have vanishing integral over
$x$ at all scales $Q^2$, because the proton
does not carry the charm quantum number, but the PDF itself may well
be nonzero, as it happens for the strange valence PDF $s^-=s-\bar s$.
Indeed,  a nonvanishing charm valence component 
is always  generated, like
for any other quark flavor, by perturbative QCD evolution~\cite{Catani:2004nc}.
However, any perturbatively generated valence charm component is tiny
in comparison to all other PDFs, including those of heavy quarks. Hence,  any
evidence of a sizable valence charm PDF is  a definite sign of its intrinsic
nature.
Model
calculations~\cite{Brodsky:2015fna,Hobbs:2013bia}, while in
broad agreement on the shape of  total IC PDF,
widely differ in   predictions for the shape and magnitude of the
intrinsic valence charm component.
Model calculations of IC complemented with input from lattice QCD~\cite{Sufian:2020coz}
also predict a non-vanishing valence component.

Here we investigate this issue by performing a data-driven
determination of the intrinsic valence charm PDF of the proton, based on
the same methodology as in~\cite{Ball:2022qks}.
We generalize the NNPDF4.0 PDF determination by introducing an
independent parametrization of the charm
and anticharm PDFs, determine them from a global QCD analysis,
and subtract the perturbatively generated contributions by
transforming all PDFs to the three-flavor-number scheme (3FNS) in
which perturbative charm vanishes so any residual charm PDF is
intrinsic.

We find a non-zero charm valence PDF, with a positive valence peak for $x\sim 0.3$,
whose local significance is close to  two sigma. 
We demonstrate the stability of this result with respect to theoretical, dataset, and 
methodological variations.
We then  propose two novel experimental probes to further scrutinize
  this asymmetry between charm and anticharm PDFs: $D$-meson
  asymmetries in $Z$+c-jet production at
  LHCb~\cite{Boettcher:2015sqn,LHCb:2021stx}, and
  flavor-tagged structure functions at the upcoming
  Electron-Ion Collider (EIC)~\cite{Kelsey:2021gpk,AbdulKhalek:2021gbh}.\\[-0.3cm]

\noindent
{\bf Methodology.}
As in Ref.~\cite{Ball:2022qks}, we follow the NNPDF4.0 methodology,
theory settings and dataset~\cite{NNPDF:2021njg}, the only
modifications being related to the independent parametrization of the
charm valence PDF. Firstly,  the neural network architecture is extended 
with an additional neuron in the output layer in order
to independently parametrize $c^-(x,Q_0)$,  Eq.~(\ref{eq:cmin}), at the PDF
parametrization scale $Q_0=1.65$~GeV.
In the default PDF  basis (``evolution basis'', see App.~\ref{app:basis_dependence})
this extra neuron is taken to
parametrize the 
valence non-singlet
combination $V_{15}=(u^- + d^- + s^- - 3c^-)$, with
$q^- \equiv q-\bar{q}$. In an  alternative basis (``flavor basis'')
it instead parametrizes  $\bar{c}$: so in both cases the valence
component is obtained by taking linear combinations of the neural
network outputs.
In 
our previous analysis~\cite{NNPDF:2021njg}, the assumption of vanishing
intrinsic valence was enforced by setting  $V_{15}=V=\sum_i q_i^-$ in the
evolution basis or $\bar{c}=c$ in the flavor basis at the scale $Q_0$.

In addition to experimental constraints, a non-zero charm
valence must, as mentioned, satisfy the sum rule
\begin{align}
\label{eq:vsr}
Q_{15}\equiv \int_0^1 dx\, V_{15}(x,Q_0)&=3 \, ,
\\ \label{eq:vsra} Q_c\equiv \int_0^1 dx\, (c-\bar{c})(x,Q_0)&=0  \,,
\end{align}
in the evolution or flavor basis respectively.
This sum rule is enforced in the same manner as that of the strange
valence sum rule~\cite{NNPDF:2021njg}.
Finally,  to ensure cross-section positivity (at $Q^2=5$ GeV$^2$)
 separately for charm- and anticharm-initiated processes, we replace the neutral current $F_2^c$ positivity
observable (sensitive only to $c^+$) with
its charged current-counterparts $F_2^{c,W^-}$ and $F_2^{\bar{c},W^+}$.
The charm PDFs $xc$ and $x\bar{c}$ themselves are not required to be
positive-definite~\cite{Candido:2020yat,Collins:2021vke,Candido:2023ujx}.
Integrability and preprocessing are imposed as in NNPDF4.0.
We have verified that results are stable upon repeating 
the hyperoptimization of all
parameters defining the fitting algorithm, and thus we keep the same
settings as in~\cite{NNPDF:2021njg}. 
\\[-0.3cm]

\noindent
{\bf The valence charm PDF.}
As explained in Ref.~\cite{Ball:2022qks}, intrinsic charm is the charm
PDF in the 3FNS, where charm is treated as a massive particle that does not
contribute to the running of the strong coupling or the evolution of
PDFs.
In the absence of intrinsic charm (``perturbative charm'', henceforth),
the charm and anticharm PDFs in the 
3FNS vanish identically.
In the four-flavor-number scheme (4FNS), in which charm is
treated as a massless parton, these PDFs are determined by
perturbative matching conditions between the 3FNS and the
4FNS~\cite{Buza:1997mg}.
In our data-driven approach, 
the charm and anticharm PDFs, instead of being fixed by perturbative
matching conditions, are determined from
data on the same footing as the light quark PDFs.
The  deviation of
data-driven charm from perturbative charm, i.e., in the 3FNS the
deviation of the charm and anticharm PDFs from zero,  is identified with the
intrinsic component. In practice, we parametrize PDFs at
$Q_0=1.65$~GeV in the 4FNS, and then invert the matching conditions to
determine the intrinsic component in the 3FNS.

\begin{figure}[t]
    \begin{center}
    \makebox{\includegraphics[width=0.99\columnwidth]{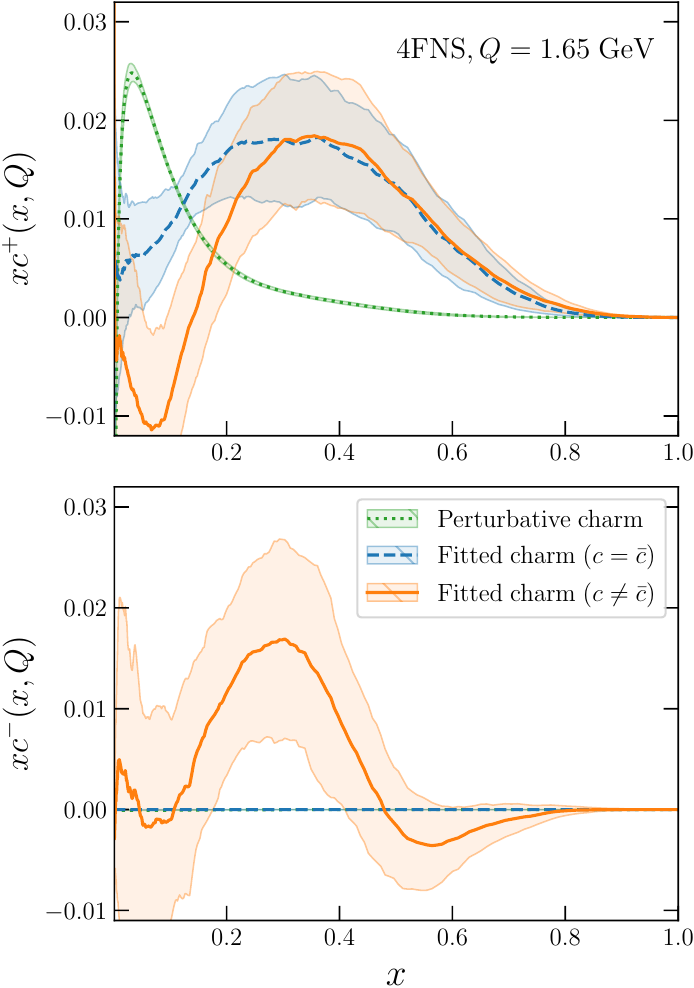}}
    \end{center}
  \vspace{-0.55cm}
  \caption{The charm total  $xc^+$ (top) and valence $xc^-$
    (bottom) PDFs in the 4FNS at $Q=1.65$ GeV. The perturbative and
    data-driven results are compared, in the latter case 
    either assuming $c^-=0$ (as in~\cite{Ball:2022qks}) or
    $c^-$ determined from data.
  }
  \label{fig:CharmAsymmetry-q1p65gev-Fit1Main}
\end{figure}

In  Fig.~\ref{fig:CharmAsymmetry-q1p65gev-Fit1Main}
we show $xc^+$ and $xc^-$ in the 4FNS at
$Q=1.65$~GeV, i.e.\ just above the charm mass that we take to be
$m_c=1.51$~GeV, determined using next-to-next-to
leading order (NNLO) QCD theory. The bands are  68\% confidence level (CL) PDF uncertainties.
 We show both the purely perturbative and data-driven results,
in the latter case both for $c=\bar{c}$
(same as in~\cite{Ball:2022qks})
and  $c \ne \bar{c}$. Note that the purely perturbative valence PDF
vanishes at $Q=m_c$ at NNLO, and only develops a tiny component at one
extra perturbative order (N$^3$LO), or at higher scales. Hence, a
nonvanishing valence component in the 4FNS provides already evidence for
intrinsic charm.

Upon allowing for a vanishing  valence  $xc^-$ component, the total
charm $xc^+$ is quite stable, especially around the peak at
$x\sim0.4$. This total charm PDF is also somewhat suppressed for
smaller $x\lesssim 0.2$ as compared to the baseline result.
In terms of fit quality, the $\chi^2$ per data point for the global
dataset decreases from 1.162 to 1.151, corresponding
to an improvement by about 50 units in absolute $\chi^2$. The
main contributions to this decrease comes from  neutral current
deep-inelastic scattering and LHC gauge boson production data (see App.~\ref{sec:fit_quality}).

The valence component is nonzero  and positive at more than one sigma level
in the  $x \in [0.2, 0.4]$ region, and consistent with zero
within the large PDF uncertainties elsewhere.
The size and shape of the valence charm PDF seen in
Fig.~\ref{fig:CharmAsymmetry-q1p65gev-Fit1Main} are stable upon
variations of PDF parametrization basis
(App.~\ref{app:basis_dependence}), the value  of $m_c$,
(App.~\ref{app:mc_variations}), the
input dataset (App.~\ref{app:dataset_dependence}),
and the kinematic cuts in $W^2$ and $Q^2$ (App.~\ref{sec:stability_cuts}).
All other PDFs are  mostly left unaffected by having allowed for a nonvanishing
valence charm.

  \begin{figure}[t]
    \begin{center}
    \makebox{\includegraphics[width=0.99\columnwidth]{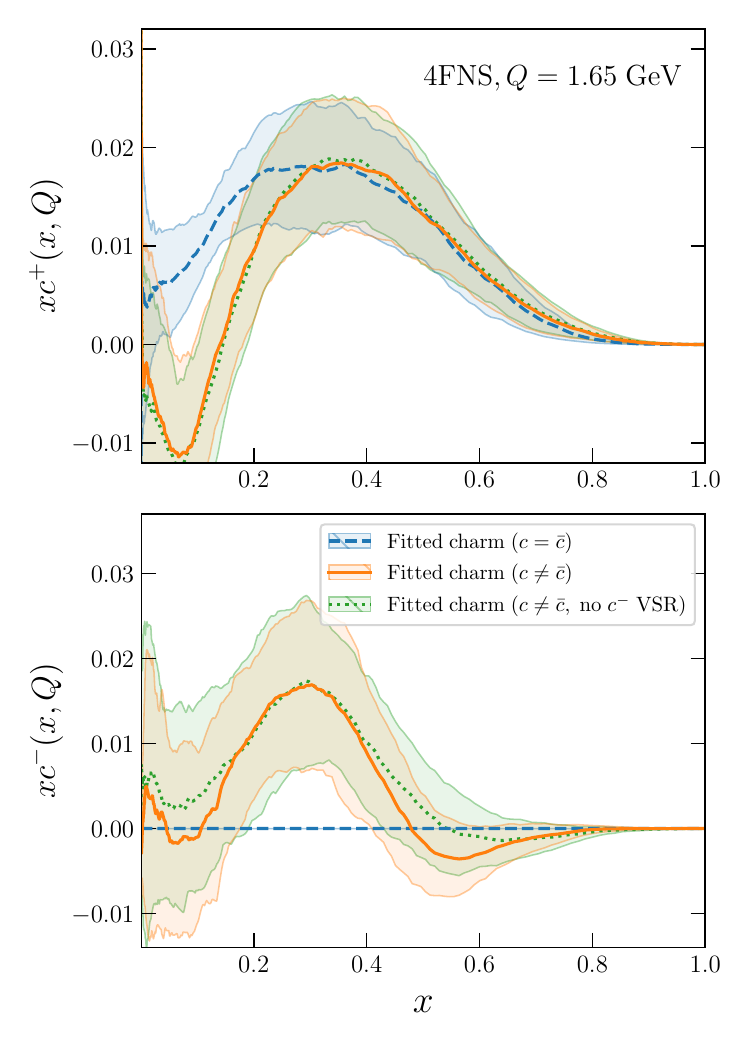}}
    \end{center}
  \vspace{-0.55cm}
  \caption{Same as Fig.~\ref{fig:CharmAsymmetry-q1p65gev-Fit1nsr}, 
    now without imposing the charm valence sum rule Eq.~(\ref{eq:vsr})
    when  $c\not=\bar c$.}
  \label{fig:CharmAsymmetry-q1p65gev-Fit1nsr}
\end{figure}

Whereas in our default determination we have imposed the charm valence sum rule
Eq.~(\ref{eq:vsr}), we have also repeated our determination without
imposing this theoretical constraint.
We then obtain $Q_c=0.07\pm 0.14$ and the resulting charm PDFs are
shown in Fig.~\ref{fig:CharmAsymmetry-q1p65gev-Fit1nsr}.
This result demonstrates
that the valence sum rule is actually enforced by the data, and
our result is data-driven.
\\[-0.3cm]

\begin{figure}[t]
    \begin{center}
    \makebox{\includegraphics[width=0.99\columnwidth]{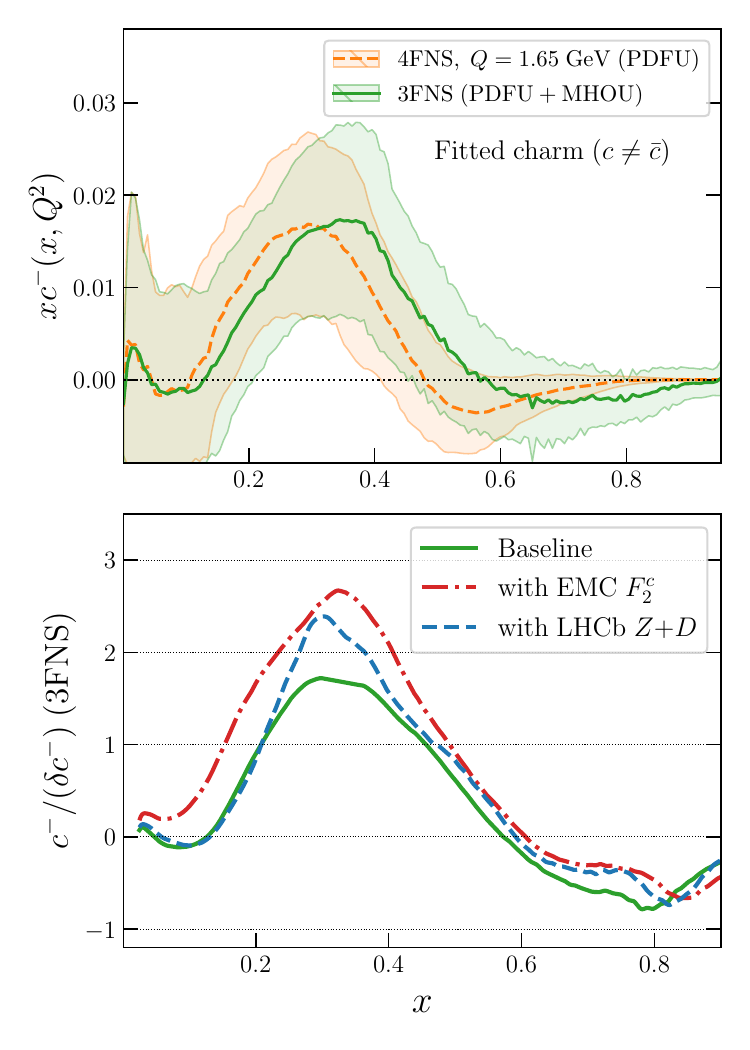}}
    \end{center}
  \vspace{-0.55cm}
  \caption{Top: the 3FNS (intrinsic) valence charm PDF  $xc^-$, compared
    to the 4FNS result (same as
    Fig.~\ref{fig:CharmAsymmetry-q1p65gev-Fit1Main} bottom).
    The 3FNS also includes MHOU due to the inversion from the 4FNS to the 3FNS.
    Bottom: the pull for  valence $xc^-$ charm PDF
    in the 3FNS.
    Results are shown both for the default fit and also when including the
 EMC $F_2^c$ and  LHCb $Z+c$ data}.
  \label{fig:3FNS_ICasy}
\end{figure}

\noindent
{\bf Intrinsic valence charm.}
The intrinsic valence charm PDF is now determined by transforming back
to the 3FNS scheme, and  is displayed in
Fig.~\ref{fig:3FNS_ICasy} (upper panel), together with its 4FNS counterpart already
shown in Fig.~\ref{fig:CharmAsymmetry-q1p65gev-Fit1Main}. An estimate
of the missing higher order uncertainties (MHOU) related
to the truncation of the perturbative expansion is also
included.
This, as in~\cite{Ball:2022qks}, is estimated  as the
change in the 3FNS PDF when
the transformation from the 4FNS to the
3FNS is performed to one higher perturbative order, i.e.\ N$^3$LO~\cite{Bierenbaum:2009zt,Bierenbaum:2009mv,Ablinger:2010ty,Ablinger:2014vwa,Ablinger:2014uka,Behring:2014eya,Ablinger_2014,Ablinger:2014nga,Blumlein:2017wxd}, as
this is estimated to be the dominant missing higher order correction.

The 3FNS and 4FNS valence PDFs turn out to be  quite close, implying
that for the valence PDF,
unlike for the total charm PDF, the theory uncertainty is smaller than
the PDF uncertainty. 
We thus find that the intrinsic (3FNS) charm valence is nonzero and positive
roughly in the same $x$ region as its 4FNS counterpart.

The statistical significance of the nonvanishing valence is quantified
by the pull, defined as the median PDF in units of the total uncertainty,
shown in Fig.~\ref{fig:3FNS_ICasy} (bottom).
The local significance of the intrinsic valence is slightly below two sigma,
peaking at $x\sim 0.3$. The significance of the total intrinsic
component is similar to that found in Ref.~\cite{Ball:2022qks}, namely
about three sigma for $x\sim 0.5$. As in Ref.~\cite{Ball:2022qks}, we also
show the results found in fit variants including 
the EMC $F_2^c$~\cite{Aubert:1982tt}  and LHCb $Z+c$ data~\cite{LHCb:2021stx}, both of which
increase the local significance.

The results of Figs.~\ref{fig:CharmAsymmetry-q1p65gev-Fit1Main}-\ref{fig:3FNS_ICasy}
suggest that the intrinsic valence component may be nonzero, 
but their significance falls below the three sigma evidence level.
We thus propose two novel experimental observables engineered to probe
this valence charm component.\\[-0.3cm]

\noindent
{\bf Charm asymmetries in $Z+c$ at LHCb.}
The LHCb LHC Run~2 data, which, as shown in Ref.~\cite{Ball:2022qks}, reinforce the evidence
for an intrinsic total charm component, correspond to
measurements of forward $Z$
production in association with charm-tagged jets~\cite{LHCb:2021stx}. 
They are presented as a measurement of $\mathcal{R}_j^c(y_Z)$, the ratio
between $c$-tagged and untagged jets in bins of the $Z$-boson rapidity
$y_Z$, and they are obtained from tagging   
$D$-mesons from displaced vertices.
The higher statistics available first at Runs~3 and 4 and later at the HL-LHC will enable
the reconstruction of the exclusive decays of $D$-mesons, and thus the separation of charm and  anticharm-tagged final states. 
We thus define the asymmetry
\be
\label{eq:charm_lhcb_asym}
\mathcal{A}_c(y_Z) \equiv \frac{N_j^c(y_Z) - N_j^{\bar{c}}(y_Z) }{
N_j^c(y_Z) + N_j^{\bar{c}}(y_Z) } \, ,
\ee
where $N_j^c$~($N_j^{\bar{c}}$) is defined in the same manner as $\mathcal{R}_j^c$~\cite{LHCb:2021stx},
but now restricted to events with $D$-mesons containing a charm quark (antiquark).
This asymmetry is directly sensitive to a possible difference between the charm
and anticharm PDFs in the initial state.

\begin{figure}[t]
    \begin{center}
    \makebox{\includegraphics[width=0.99\columnwidth]{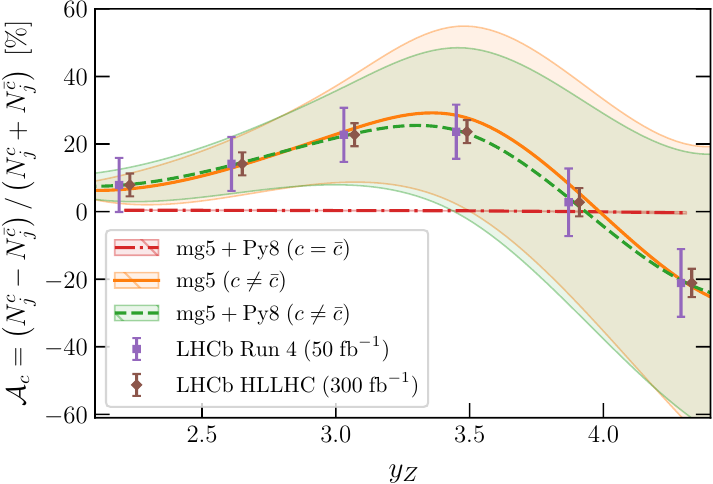}}
    \end{center}
  \vspace{-0.55cm}
  \caption{
    The charm asymmetry $\mathcal{A}_c(y_Z)$, Eq.~(\ref{eq:charm_lhcb_asym}), in $Z$+$c$-jet production at LHCb ($\sqrt{s}=13$ TeV)
    evaluated at LO matched to parton showers with the nonvanishing valence PDF determined here.
    The pure LO result and the result with vanishing charm valence are
    also shown for comparison.
    The bands correspond to one-sigma PDF uncertainties.
    Projected  statistical uncertainties for LHCb measurements at Run~4 ($\mathcal{L}=50~$fb$^{-1}$) and the HL-LHC ($\mathcal{L}=300~$fb$^{-1}$) are also shown.}
  \label{fig:LHCb_zcham_HLLHC}
\end{figure}

In Fig.~\ref{fig:LHCb_zcham_HLLHC} we display the asymmetry $\mathcal{A}_c(y_Z)$, Eq.~(\ref{eq:charm_lhcb_asym}),
computed for  $\sqrt{s}=13$~TeV using the PDFs determined here, that
allow for a nonvanishing valence component, as well as the default 
NNPDF4.0 with $c=\bar{c}$.  Results are computed
using {\sc\small mg5\_aMC@NLO}~\cite{Alwall:2014hca} at leading order (LO)
matched to {\sc\small Pythia8}~\cite{Sjostrand:2007gs,Skands:2014pea},
with the same $D$-meson tagging and jet-reconstruction
algorithm as in~\cite{Boettcher:2015sqn,LHCb:2021stx}.
The leading order  parton-level result is also shown.

It is apparent from Fig.~\ref{fig:LHCb_zcham_HLLHC} that, even though the forward-backward asymmetry of the $Z$
decay generates a small
asymmetry $\mathcal{A}_c\ne 0$ even when $c=\bar{c}$~\cite{Gauld:2015qha,Gauld:2019doc}, the LO effect due to an
asymmetry between $c$ an $\bar c$ PDFs is much larger, and 
stable upon showering and hadronization corrections. Indeed, higher-order QCD corrections largely cancel in the ratio $\mathcal{A}_c(y_Z)$.

In Fig.~\ref{fig:LHCb_zcham_HLLHC} we also display projected uncertainties
for the LHCb measurement of this asymmetry at Run~3 and at the HL-LHC
(see App.~\ref{sec:zcharm_lhcb} for details),  showing
that a valence component
of the same size as our central prediction could be detected respectively
at about a two sigma or four sigma level.
\\[-0.3cm]

\noindent
{\bf Charm-tagged DIS at the EIC.}
A standard probe of the charm component of the proton is the
deep-inelastic charm
structure function $F_2^c$~\cite{Aubert:1982tt,Forte:2010ta,Ball:2015dpa,H1:2018flt} and the
associate deep-inelastic reduced charm  production cross-section
$\sigma_{\rm red}^{c\bar{}c}$. Correspondingly, the charm valence can
be determined from the reduced cross-section asymmetry 
\be
\label{eq:EIC_asy_F2c}
\mathcal{A}_{\sigma^{c\bar{c}}}(x,Q^2) \equiv \frac{\sigma_{\rm red}^{c}(x,Q^2)- \sigma_{\rm red}^{\bar{c}}(x,Q^2)}{
\sigma_{\rm red}^{c\bar{c}}(x,Q^2)} \, .
\ee
A measurement of this observable requires reconstructing final-state  $D$-mesons by identifying
their decay products.
At the future EIC this will be possible
with good precision using the proposed ePIC detector~\cite{Khalek:2021ulf,Armesto:2023hnw,Kelsey:2021gpk}.

  \begin{figure}[t]
    \begin{center}
      \makebox{
         \includegraphics[width=0.99\columnwidth]{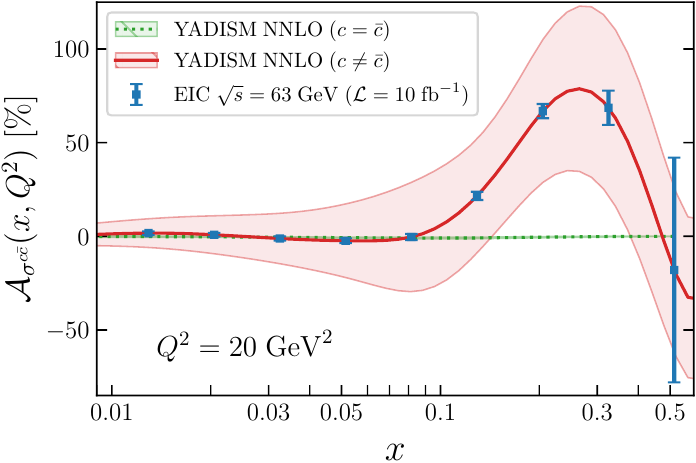}
      }
    \end{center}
  \vspace{-0.55cm}
  \caption{The  reduced charm-tagged cross-section asymmetry
    $\mathcal{A}_{\sigma^{c\bar{c}}}$, Eq.~(\ref{eq:EIC_asy_F2c}), at
    $Q^2=20$~GeV$^2$ computed at NNLO QCD using the nonvanishing
    valence PDF determined here.
    The result  with vanishing charm valence is
    also shown for comparison.
    The bands correspond to one-sigma PDF
    uncertainties.
    The projected  statistical uncertainties at the
    EIC~\cite{Kelsey:2021gpk} (running at $\sqrt{s}=63$
    GeV for $\mathcal{L}=10$ fb$^{-1}$) are also shown.
  }
  \label{fig:F2c-EIC}
\end{figure}

  The predicted asymmetry $\mathcal{A}_{\sigma^{c\bar{c}}}$ at $Q^2=20$~GeV$^2$
is  shown in Fig.~\ref{fig:F2c-EIC}; results are shown at the reduced
charm (parton) cross-section level, evaluated with {\sc\small YADISM}~\cite{Candido:2023utz}
at NNLO accuracy.
As in Fig.~\ref{fig:LHCb_zcham_HLLHC}, we show results obtained both using the PDFs determined here, that
allow for a nonvanishing valence component, as well as the default 
NNPDF4.0 with $c=\bar{c}$.
We also display the projected statistical uncertainties~\cite{Kelsey:2021gpk}  at the EIC running
at $\sqrt{s}=63$ GeV for $\mathcal{L}=10$ fb$^{-1}$ (see
App.~\ref{SI:EIC}). It is clear that a nonvanishing charm valence
component  can be measured at the EIC to very high significance even for a moderate
amount of integrated luminosity.

  \begin{figure}[t]
    \begin{center}
      \includegraphics[width=0.99\columnwidth]{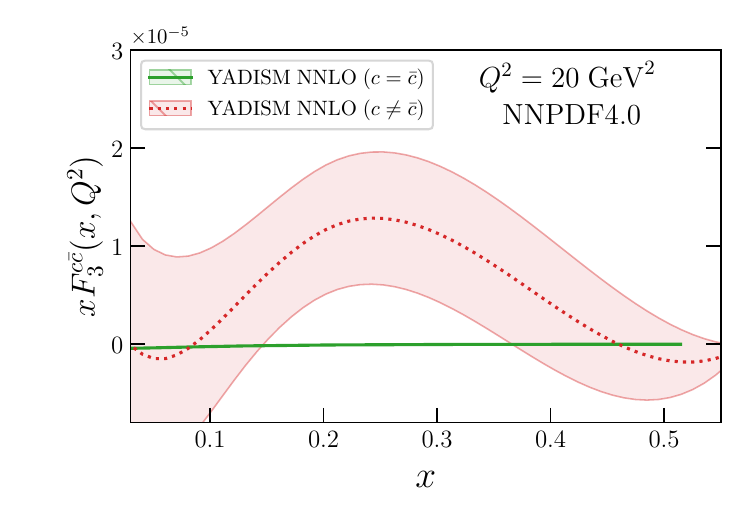}
      \end{center}
  \vspace{-0.55cm}
  \caption{Same as Fig.~\ref{fig:F2c-EIC}  for the charm-tagged parity-violating
  structure function $xF_3^{c\bar{c}}(x,Q^2)$ at the EIC (no
  projection for the statistical accuracy of the EIC mmeasurement is available).
  }
  \label{fig:ICasym-SI-EIC-xF3c}
\end{figure}

In addition to the charm-tagged  structure function $F_2^{c\bar{c}}$, at the EIC
complementary sensitivity to the charm valence content of the proton would be provided
by the charm-tagged parity-violating
  structure function $xF_3^{c\bar{c}}(x,Q^2)$.
  This observable has the advantage that at LO is already proportional
  to $xc^-$, hence  provides a direct constraint on valence charm.
  Predictions for this observable, are presented in 
Fig.~\ref{fig:ICasym-SI-EIC-xF3c}. Even in the absence of detailed 
predictions for prospective EIC
measurements of this observable, it is clear that its measurement would
significantly constrain the charm valence PDF.\\[-0.3cm]


\noindent
    {\bf Outlook.}
    Our main conclusion is that current experimental data  provide support
    for the hypothesis that the valence charm PDF may be nonzero, even
    though with the NNPDF4.0 dataset it is not possible to reach three-sigma evidence.
Whereas the
situation may improve somewhat with future PDF determinations based on
the full LHC Run-3 dataset, dedicated observables such as the LHCb
charm asymmetry Eq.~(\ref{eq:charm_lhcb_asym}) as well as charm production
at the EIC Eq.~(\ref{eq:EIC_asy_F2c}) will be needed in order
to achieve firm evidence or discovery.
Other experimental probes that could be explored in this context
include open charm production and asymmetries at the LHC, in particular for forward (LHCb~\cite{LHCb:2022cul,LHCb:2018jry})
and far-forward (FASER$\nu$~\cite{FASER:2023zcr}, SND@LHC~\cite{SNDLHC:2023pun},
and the Forward Physics
Facility~\cite{Feng:2022inv,Cruz-Martinez:2023sdv})
detectors. Progress in lattice computations might well also provide
further constraints.

From the theory point of view, ongoing efforts towards a NNPDF determination
based on N$^3$LO calculations should reduce some of the theory uncertainties
affecting the current determination.
On a more speculative vein, it
might also be interesting to investigate an intrinsic bottom quark component and its eventual
asymmetry.

\paragraph{Acknowledgments.}
We are grateful to Rhorry Gauld for extensive discussions concerning $Z$+charm production,
and to Reynier Cruz-Torres and Barak Schmookler by providing the EIC projections for $F_2^c$.
We thank Thomas Boettcher,  Dan Craik, Philip Ilten, Patrick Koppenburg, Niels Tuning,  
and Michael Williams,
for information about the LHCb $Z$+charm measurements and the associated projections.
R.~D.~B. and R.~S. are supported by the U.K.
Science and Technology Facility Council (STFC) grant ST/T000600/1. 
T.~G. is supported by NWO (Dutch Research Council) via an ENW-KLEIN-2 project.
F.~H. is supported by the Academy of Finland project 358090 and is funded as a part of the Center of Excellence in Quark Matter of the Academy
of Finland, project 346326.
E.R.~N. is supported by the Italian Ministry of University and Research (MUR)
through the “Rita Levi-Montalcini” Program.
J.~R. and G.~M. are partially supported by NWO (Dutch Research Council).

\bibliography{ICasym}

\appendix
\onecolumngrid

\vspace{1cm}
\begin{center}
{\Large \bf Appendix}
\end{center}

We collect here a number of supplementary results, 
none of which  is required
for the understanding of the main text, which is self-contained.
The purpose of this Appendix is (a) to ensure
reproducibility of our results and (b) to document the set of tests of
stability and robustness that we have carried out.
In particular: we document the fit quality for our default determination;
we test for stability upon changes of parametrization basis,
variations of the quark mass, choice of dataset and kinematic cuts; and
we provide details of our projections for measurements at the
HL-LHC and the EIC.

\section{Fit quality and data impact}
\label{sec:fit_quality}

We compare the fit quality for the PDF determination presented here
with $c\ne \bar{c}$ to the published  NNPDF4.0 determination
with $c = \bar{c}$, by showing in Table~\ref{tab:chi2-baseline}  the
experimental $\chi^2$ per data point 
for different groups of processes and for the total dataset.
We refer to~\cite{NNPDF:2021njg} for the
definition of the dataset and of the $\chi^2$ and the process
categories (see in particular Table~5.1 of Ref.~\cite{NNPDF:2021njg}).

\begin{figure}[t]
    \begin{center}
      \includegraphics[width=0.6\columnwidth]{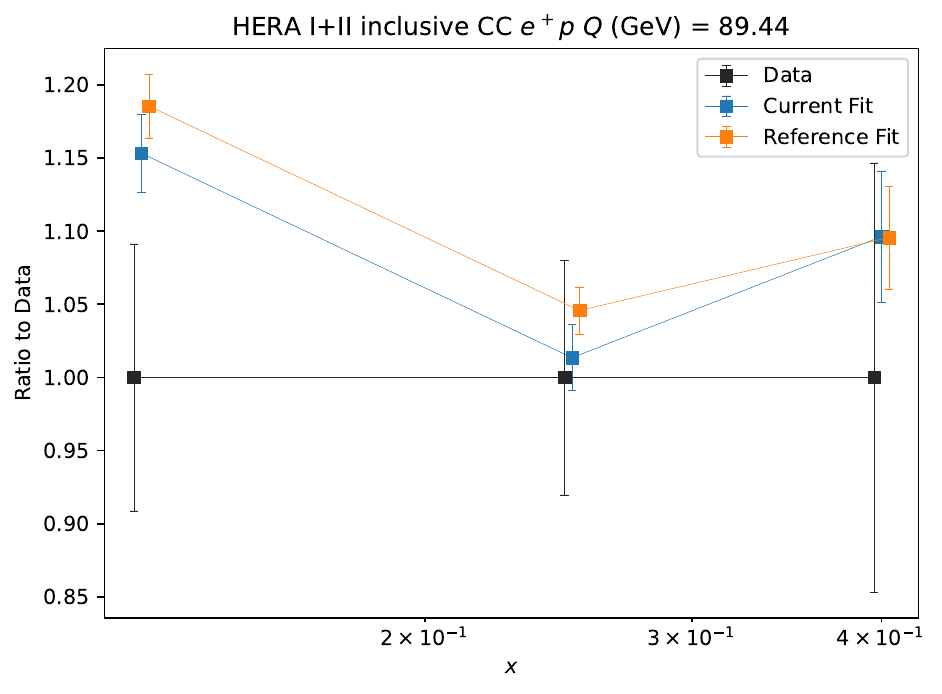}
      \end{center}
  \vspace{-0.55cm}
  \caption{Comparison between the data (black) and theory prediction obtained
    using the $c \ne \bar{c}$ (blue) and $c = \bar{c}$ (orange) PDF
    sets of Table~\ref{tab:chi2-baseline}, for HERA $e^+p$ charged-current DIS
    in one of the largest $Q^2$ bins, $Q=89.44$ GeV, shown as a ratio to the
    experimental data.
  }
  \label{fig:ICasym-SI-heracc}
\end{figure}

The largest reduction in absolute $\chi^2$ upon allowing for a
nonvanishing charm valence component is in collider DIS (i.e. HERA), and the
largest reduction in  $\chi^2$ in charged-current collider
DIS. The largest impact is seen in the large $Q^2$, large $x$ $e^+p$
CC bins
such as shown in Fig.~\ref{fig:ICasym-SI-heracc}, consistent with the
observation that the intrinsic charm PDFs are localized at large $x$.
Note that HERA data
for the
$F_2^c$ charm structure function, that are included in the fit, have
no impact on intrinsic charm because they are in the medium-to-low $x$ region
where the charm PDF is dominated by the perturbative component.

\begin{table}[!t]
\centering
\small
\renewcommand{\arraystretch}{1.90}
\begin{tabularx}{\textwidth}{Xrcc}
\hline
Dataset  &  $n_{\rm dat}$
         & $\qquad \chi^{2}/n_{\rm dat}$ ($c \ne \bar{c}$) $\qquad$
         & $\qquad\chi^{2}/n_{\rm dat}$ ($c = \bar{c}$) $\qquad$\\
\hline
DIS NC (fixed-target)                     &  973 & 1.24 & 1.26 \\
DIS CC (fixed-target)                     &  908 & 0.86 & 0.86 \\
DIS NC (collider)                         & 1127 & 1.18 & 1.19 \\
DIS CC (collider)                         &   81 & 1.23 & 1.28 \\
Drell-Yan (fixed-target)                  &  195 & 1.02 & 1.00 \\
Tevatron $W$, $Z$ production (inclusive)  &   65 & 1.06 & 1.09 \\
LHC $W$, $Z$ production (inclusive)       &  463 & 1.35 & 1.37 \\
LHC $W$, $Z$ production ($p_T$ and jets)  &  150 & 0.99 & 0.98 \\
LHC top-quark pair production             &   64 & 1.28 & 1.21 \\
LHC jet production                        &  171 & 1.25 & 1.26 \\
LHC isolated $\gamma$ production          &   53 & 0.76 & 0.77 \\
LHC single $t$ production                 &   17 & 0.36 & 0.36 \\
\hline
{\bf Total}                               & {\bf 4616} & {\bf 1.151} & {\bf 1.162}\\
\hline
\end{tabularx}

\caption{\small The values of the experimental $\chi^2$ per data point
for the different groups of processes entering the NNPDF4.0 determination as
well as for the total dataset. We compare the results of the baseline NNPDF4.0
fit ($c = \bar{c}$) with the results of this work ($c \ne \bar{c}$).}
\label{tab:chi2-baseline}
\end{table}

\section{Parametrization basis dependence}
\label{app:basis_dependence}

The NNPDF4.0  set is determined by choosing by default as a
parametrization basis for PDFs the eigenstates of QCD evolution; as a
consistency check results in Ref.~\cite{NNPDF:2021njg}
are presented by alternatively choosing
flavor and antiflavor eigenstates.
Because here we are determining a
difference between PDFs, it is especially important to check stability
upon choice of basis.
In the evolution basis, the charm PDFs are parametrized through the
two combinations
\be
T_{15}(x,Q^2) = \lp u^+ + d^+ + s^+ - 3c^+\rp(x,Q) \, , \quad V_{15}(x,Q^2) = \lp u^- + d^- + s^- - 3c^-\rp (x,Q) \,, 
\ee
at a scale $Q=Q_0=1.65$~GeV.
In the flavor basis, they are  parametrized as $c(x,Q)$ and
$\bar{c}(x,Q)$.
Note that in neither case the  total and valence combinations $c^\pm$ are elements of the basis.

  \begin{figure}[t]
    \begin{center}
      \includegraphics[width=0.99\columnwidth]{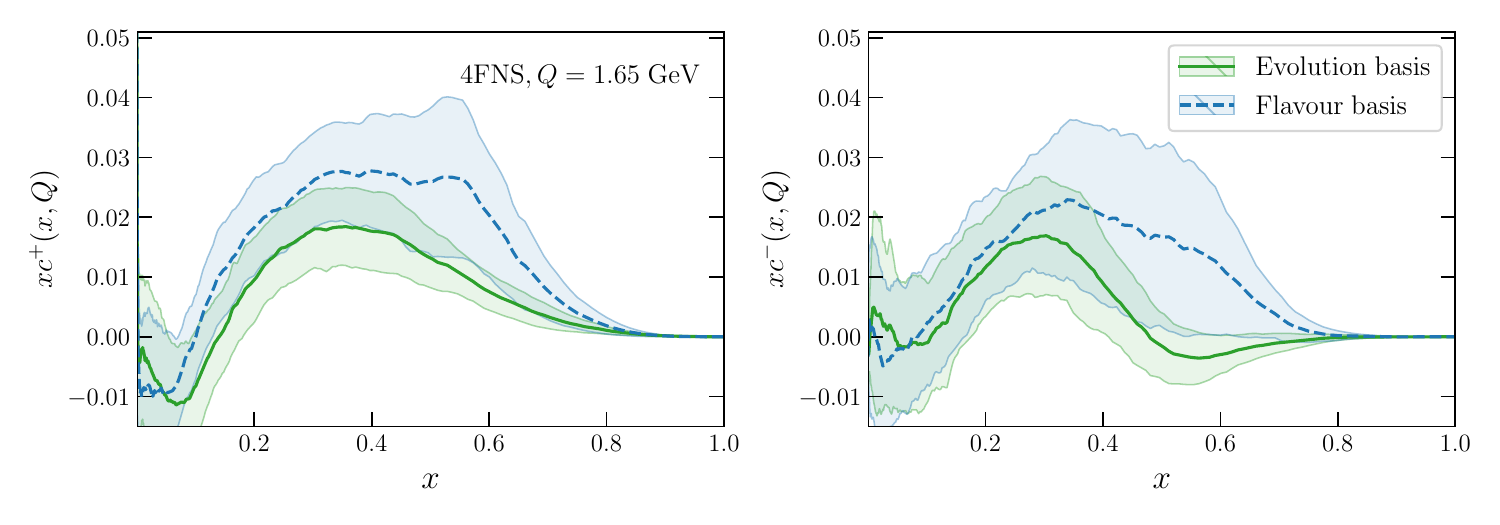}
      \end{center}
  \vspace{-0.55cm}
  \caption{Comparison between the total  $xc^+$ (left) and valence $xc^-$ (right) charm
  PDFs 
in the 4FNS at $Q=1.65$ GeV, obtained parametrizing PDFs in the  evolution basis
 (default) and in the flavor basis.
  }
  \label{fig:ICasym-SI-basisdep}
\end{figure}

In Fig.~\ref{fig:ICasym-SI-basisdep} the $xc^\pm$ PDFs found using
either basis are compared. 
Agreement at the one sigma level is found for all $x$.
The main qualitative features are independent of the basis choice,
specifically the presence of a positive valence peak around $x\sim0.3$
for $xc^-$.
 Results in the flavor basis display
larger PDF uncertainties, possibly because the flavor basis
has not undergone the same extensive hyperoptimization
as the fits in the evolution basis.

\section{Dependence on the value of the charm mass}
\label{app:mc_variations}

  \begin{figure}[t]
    \begin{center}
      \includegraphics[width=0.99\columnwidth]{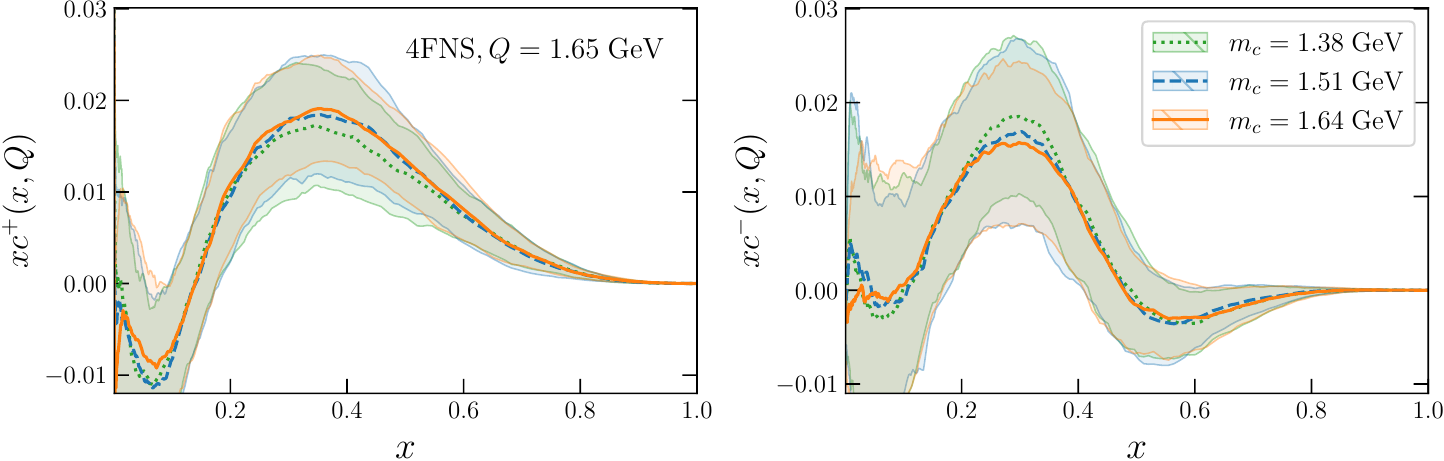}
      \end{center}
  \vspace{-0.55cm}
  \caption{Comparison of  the total  $xc^+$ (left)  and valence charm
    $xc^-$ (right) PDFs 
in the 4FNS at $Q=1.65$~GeV  as the charm pole mass is varied about
    the default central value $m_c=1.51$~GeV by $\pm\,0.13$~GeV.
  }
  \label{fig:ICasym-SI-mcdep}
\end{figure}
We verify the independence of our results on the value of the charm
quark mass, by repeating our determination as the  (pole) charm mass
is varied from the default $m_c=1.51$~GeV to $m_c=1.38$~GeV
and 1.64~GeV.
Note that we always choose the scale  $\mu_c=m_c$ as matching scale 
between the 4FNS and 3FNS, hence this is also varied  alongside $m_c$.
The total and valence charm PDFs $xc^\pm$ at  $Q=1.65$~GeV in the 4FNS
are displayed in Fig.~\ref{fig:ICasym-SI-mcdep}.
The result is found
to be essentially independent of the charm mass value, in agreement
with the corresponding result of Ref.~\cite{NNPDF:2021njg} (see Fig~8.6)

\section{Dataset dependence}
\label{app:dataset_dependence}

  \begin{figure}[t]
    \begin{center}
      \includegraphics[width=0.90\columnwidth]{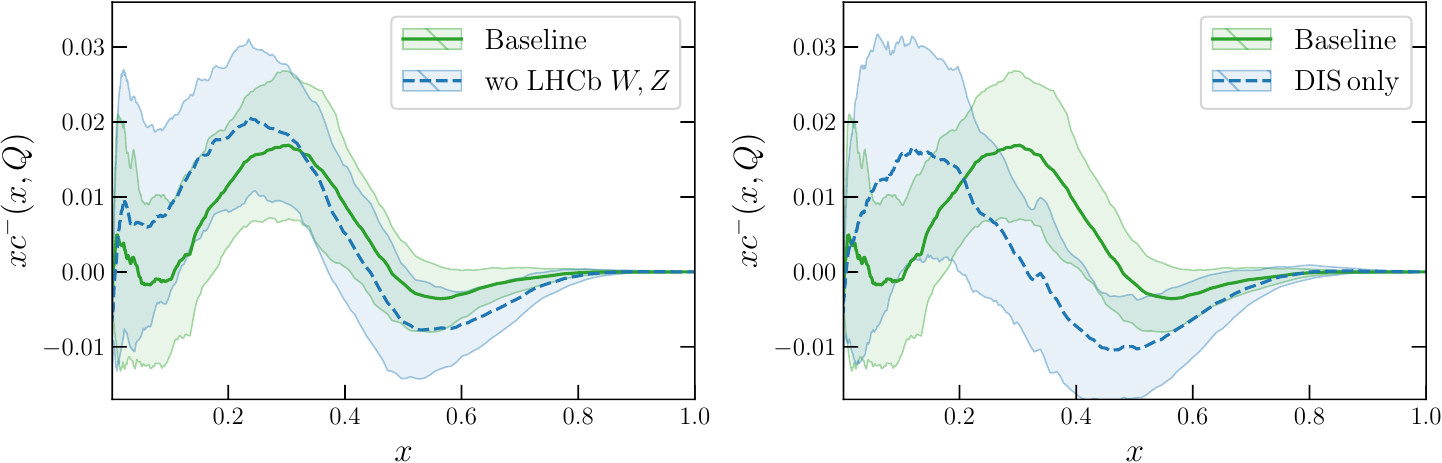}
      \end{center}
  \vspace{-0.55cm}
  \caption{The charm valence PDF in the default determination compared to a
  determination in which LHCb $W,Z$ inclusive production data are excluded (left)
and a determination based on DIS structure functions only (right).
}
  \label{fig:ICasym-SI-dataset}
\end{figure}

We study the effect of removing some datasets from our determination, with
the dual goal of checking the stability of our results, and
investigating which data mostly determine the valence charm PDF.
Specifically, we  remove the LHCb $W,Z$ data, which was found
in~\cite{Ball:2022qks} 
to dominate the constraints on the total charm PDF from all collider
measurements, and we determine the PDFs only using  DIS structure
function data.
The valence charm PDF found in either case is compared
to the default in Fig.~\ref{fig:ICasym-SI-dataset}.
Removing the LHCb electroweak data leaves $xc^-$ mostly unchanged,
hence the valence PDF appears to be less sensitive to this data than
the total charm.
When only including DIS data a nonzero valence
component is still found but now with a reduced significance: the result
is consistent with zero at the one sigma level. 

 \section{Kinematic cuts}
 \label{sec:stability_cuts}

\begin{figure}[t]
    \begin{center}
      \includegraphics[width=0.90\textwidth]{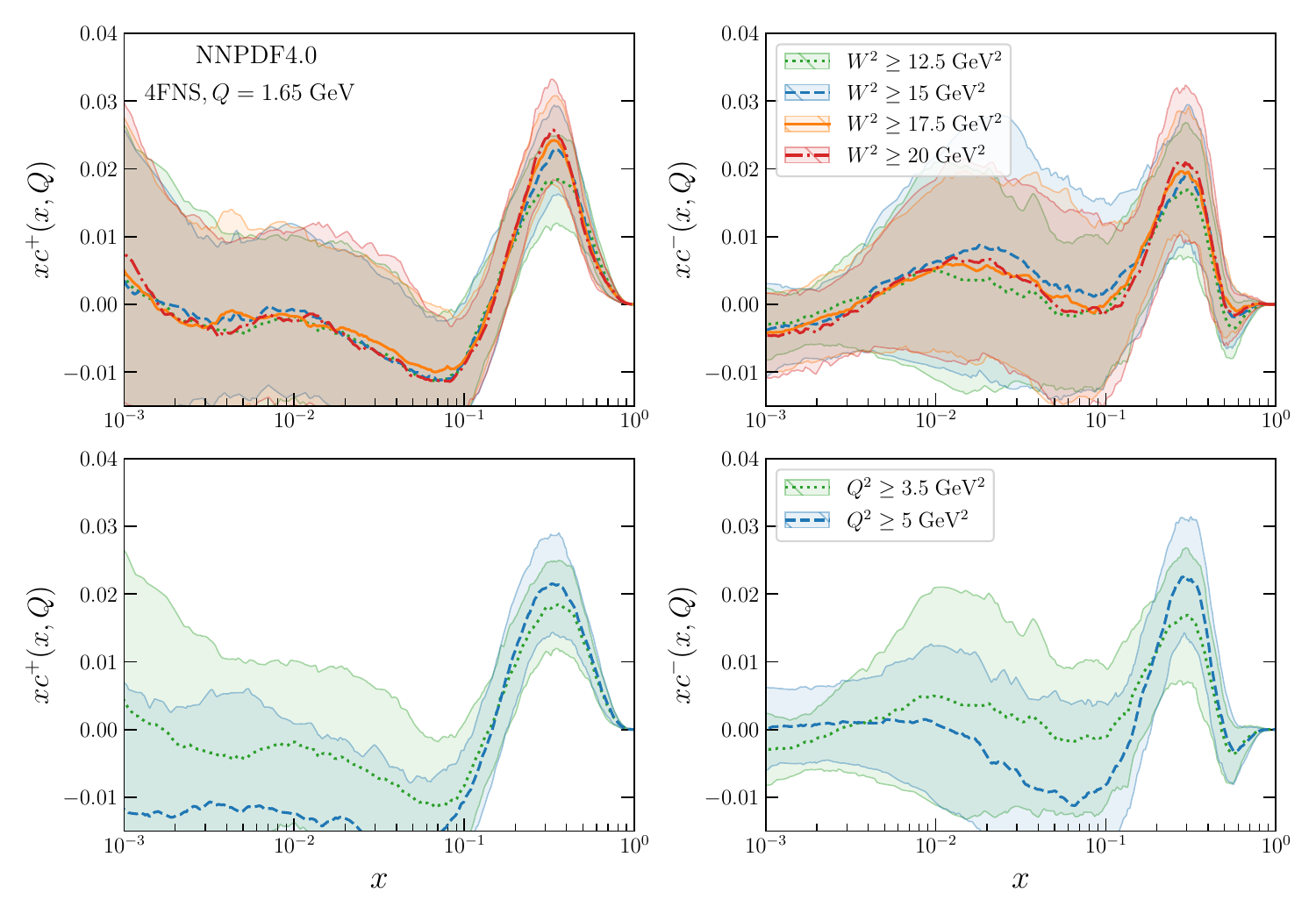}
      \end{center}
  \vspace{-0.55cm}
  \caption{The variation in the 4FNS total (left) and valence (right) charm
    PDFs at $Q=1.65$ GeV as the $W^2$ cut is raised 
to 20~GeV$^2$ in steps of 2.5~GeV$^2$ (top) and the $Q^2$ cut is
    raised to 5~GeV$^2$ (bottom).
    The kinematic cuts in the baseline fit are $Q^2\ge 3.5$ GeV$^2$
    and $W^2\ge 12.5$ GeV$^2$.
  }
  \label{fig:ICasym-SI-KinCuts}
\end{figure}

The NNPDF4.0 dataset only includes data with  $Q^2 \ge 3.5$ GeV$^2$
and $W^2 \ge 12.5$ GeV$^2$,  in order to ensure the reliability of the
leading-twist, 
fixed-order perturbative approximation.
It is important to verify that results for intrinsic charm are stable
upon variation of these cuts, as this checks that intrinsic charm is
not contaminated by possible nonperturbative corrections not accounted
for in the global PDF fitting framework.
To this purpose, we have raised the $W^2$ cut in steps of 2.5~GeV$^2$ up
to 20~GeV$^2$, and the $Q^2$ cut up to 5~GeV$^2$.
Results are displayed in
Fig.~\ref{fig:ICasym-SI-KinCuts}, and prove satisfactory stability: upon
variation of the $W^2$ cut nothing changes, and upon variation of the
$Q^2$ cut (which removes a sizable amount of data) the
central value is stable and the uncertainty only marginally increased.

\section{The $Z+$charm asymmetry at LHCb}
 \label{sec:zcharm_lhcb}

The projected statistical uncertainties for the future LHCb measurement
of $\mathcal{A}_c$ shown in Fig.~\ref{fig:LHCb_zcham_HLLHC}
are obtained extrapolating from those of the Run~2
data by correcting both for the higher luminosity and for the acceptance
associated to the different charm-tagging procedure required in this case.
The uncertainties obtained in the Run~2 measurement~\cite{LHCb:2021stx}
and based on an integrated luminosity of $\mathcal{L}=6$ fb$^{-1}$ are rescaled
to the expected luminosity to be accumulated by LHCb by the end of Run~4, $\mathcal{L} \sim 50$ fb$^{-1}$,
and at the HL-LHC, $\mathcal{L} \sim 300$ fb$^{-1}$.
Furthermore, the Run~2 measurement was based on charm-meson tagging
with displaced vertices, with a charm-tagging efficiency of $\epsilon_c \sim 25\%$.
The asymmetry $\mathcal{A}_c$ requires separating charm from anticharm in the final states,
which in turn demands reconstructing the $D$-meson decay products.
The associated efficiency is estimated by weighting the $D$-meson  branching fractions
to the occurrence of each species in the LHCb $Z$+charm sample, resulting
in an efficiency of $\epsilon_c \sim 3\%$.
The uncertainty on the asymmetry is then determined by using error
propagation with $N_j^c=N^{\bar{c}}_j$, neglecting the dependence of the
uncertainty on the value of the asymmetry itself.

\section{Charm structure functions at  the Electron-Ion Collider}
\label{SI:EIC}

The projected statistical uncertainties for the future EIC  measurement
of the charm-tagged asymmetry $\mathcal{A}_{\sigma^{c\bar{c}}}$ shown in Fig.~\ref{fig:F2c-EIC}
are obtained as follows.
We adopt the projections from~\cite{Kelsey:2021gpk} for the kinematic
coverage in the  $(x,Q^2)$ plane
and the expected statistical precision, based on running at a
center-of-mass energy of  $\sqrt{s}=63$ GeV for $\mathcal{L}=10~{\rm fb}^{-1}$.
These projections entail that  measurements
of charm production at the EIC will cover the region 
$1.3~{\rm GeV}^2\lesssim Q^2\lesssim 120$ GeV$^2$ and $5\times 10^{-4}\lesssim
x\lesssim 0.5$.
Charm production is tagged from  the reconstruction of
$D^0$ and $\bar{D}^0$ exclusive decays, and a detailed estimate of
experimental uncertainties would require a full detector
simulation. Here, however, we limit ourselves to estimating the
statistical accuracy on the asymmetry Eq.~(\ref{eq:EIC_asy_F2c}), which
is expressed in terms of reduced cross-sections, defined as in
Ref.~\cite{H1:2018flt} in terms of charm structure functions.
For this, we take
the statistical uncertainties provided in~\cite{Kelsey:2021gpk} and
increase them by a factor $\sqrt{2}$ since
the measured sample has to be separated
into $D$- and $\bar{D}$-tagged subsamples.


\end{document}